\documentclass{moriond}

\def\be{\begin{equation}}
\def\ee{\end{equation}}
\def\bea{\begin{eqnarray}}
\def\eea{\end{eqnarray}}

\newcommand{\Photo}{}

\begin{document}
\vspace*{4cm}
\title{Estimating the binary neutron star merger rate density evolution with Einstein Telescope}

\author{ Neha Singh$^1$, Tomasz Bulik$^2$, Aleksandra Olejak$^3$ }

\address{1. Departament de F{\'i}sica, Universitat de les Illes Balears, IAC3–IEEC, E-07122 Palma, Spain. \\ 2. Astronomical Observatory, University of Warsaw, Al. Ujazdowskie 4, 00-478 Warsaw, Poland. \\ 3. Max Planck Institute for Astrophysics, Karl-Schwarzschild-Straße 1, 85748 Garching b. M{\"u}nchen, Germany}

\maketitle\abstracts{The Einstein Telescope (ET) is a proposed third-generation, wide-band gravitational wave (GW) detector which will have an improved detection sensitivity in low frequencies, leading to a longer observation time in the detection band and higher detection rate for binary neutron stars (BNSs). Despite the fact that ET will have a higher detection rate, a large fraction of BNSs will remain undetectable. We present a scheme to estimate accurate detection efficiency and to reconstruct the true merger rate density of the population of the BNSs, as a function of redshift. We show that with ET as a single instrumnet, for a population of BNSs with $R_{mer} \sim 100 (300)$ $\rm Gpc^{-3} yr^{-1}$ at $z\sim 0(2)$, we can reconstruct the merger rate density uptil $z \sim 2$ , with a relative error of $12\%$ at ($z \sim 2$), despite the loss in detection of the bulk of the BNS population.}

\section{Introduction}

The first gravitational wave detection of a binary neutron star (BNS) inspiral GW170817~\cite{PhysRevLett.119.161101} has provided immense momentum to modeling the evolution of these binaries. Earlier such modeling was done largely based on theoretical models, aided by a few radio and X-ray observations. The next confident BNS detection was GW190425~\cite{2020ApJ...892L...3A}. GWTC-3~\cite{2023PhRvX..13d1039A,2023PhRvX..13a1048A} detections have already provided constrains on the local merger rates of BNSs with just two events, assuming that the merger rate is constant in comoving volume out to a redshift of $z = 0.15$. A lot of progress has been made in the past decade to improve the understand of different evolution channels for the formation of the compact binaries (see Mandel \& Broekgaarden~\cite{2022LRR....25....1M} for details), leading to improved merger rate predictions for compact binaries such as binary black hole BBH, neutron star-black hole binaries (NS-BH), and BNS. In case of binaries evolving in isolation, there are still multiple sources of uncertainties such as metallicity evolution over redshift, the star formation rate across cosmic history, the dynamics of mass transfer in the binary system, and supernova engine, to name a few. These can lead to orders of magnitude variation in the merger rate predictions e.g the uncertainties in the distribution of initial conditions such as the masses of the two companions, their separations, and the binary eccentricity, can affect the expected local merger rate by a factor of upto $\sim 6$. So a strong constraint on the merger rate density evolution over redshift will, be key in limiting these uncertainties.

The next generation gravitational wave detectors such as Einstein Telescope~\cite{2011CQGra..28i4013H,2025arXiv250312263A} (ET) and Cosmic Explorer~\cite{2019BAAS...51g..35R} (CE) will be able to detect binaries with a total mass of $\sim 20 M_{\odot}$ upto $z\sim 100$, and therefore will play a pivotal role in improving the understanding of binary evolution. ET will be able to detect equal mass, optimally located BNSs of total mass $\sim 3 M_{\odot}$ up to $z\sim 2-3$. The detection rate for any GW detector is a sensitive function of total mass, mass ratio, redshift, sky-location and orientation with respect to the detector, of the binary systems. So even though, next generation detectors will have a much larger redshift reach as compared to current detectors, a large fraction of BNSs signals will not be loud enough~\cite{2012PhRvD..86l2001R,2016PhRvD..93b4018M} to cross the signal to noise detection threshold. The exact fraction detected, i.e the detection efficiency, will depend on the chosen detection threshold. In order to accurately estimate the merger rate density as a function of redshift, the \emph{detection efficiency} as a function of redshift, should be modeled precisely. In this current work we use the population-independent method to reconstruct detection efficiency and then the merger rate density developed in Singh {\it et al}~\cite{2024A&A...681A..56S}, to estimate the evolution of the merger rate density of BNSs population originating from population I and II (Pop I+II) stars.

\section{Method}

In the current work we simulate a mock population of BNSs, which are assumed to have evolved through isolated evolutionary channel. We use the M30B model from Belczynski {\it et al}~\cite{2020A&A...636A.104B} to generate our mock population. The local merger rate density prediction of this model is consistent with the GWTC-3 BNS merger rate constraints. The only change we implemented in generating this model is that we use a finer metallicity grid: 

Z = 0.00010, 0.00012, 0.00014, 0.00017, 0.00020, 0.00024, 0.00029, 0.00035, 0.00042, 0.00050, 0.00059, 0.00071, 0.00085, 0.00100, 0.00101, 0.00121, 0.00145, 0.00173, 0.00196, 0.00200, 0.00207, 0.00247, 0.00353, 0.00422, 0.00500, 0.00505, 0.00603, 0.00721, 0.00861, 0.01000, 0.01030, 0.01230, 0.01471, 0.01757, 0.02000, 0.02100, 0.02510, 0.03000.

We estimate the chirp mass and the redshift of the merging BNSs detectable with ET, using the exact threshold and method used in Singh {\it et al}~\cite{2024A&A...681A..56S}.  For the generated population of compact binaries, the true values of the chirp mass, total mass, and redshift of these `sources' are represented as $\mathcal{M}_{\rm s,mock}$, $M_{\rm s, mock}$, and $z_{\rm s, mock}$, respectively. A binary source is considered as `detected' if it crosses a detection threshold set on the signal to noise ratio (SNR). The chirp mass, total mass, and redshift of these detected sources are denoted $\mathcal{M}_{\rm s,det}$, $M_{\rm s,det}$, and $z_{\rm s, det}$, respectively. The posterior probability distribution for the chirp mass and redshift for the detected source is estimated and the median values of the estimated posterior probability distributions of chirp mass, total mass, and redshift for the detected compact binary source is represented as $\mathcal{M}_{\rm med,det}$, $M_{\rm med,det}$, and $z_{\rm med, det}$, respectively. In our mock population of BNSs, we generated a set of $10^6$ binaries, which corresponds to an observation time of 7.69 yrs. Out of these million binaries only 88681 BNSs crossed the detection threshold. 

\begin{figure}[h!]
	\begin{minipage}{0.50\linewidth}
		\centerline{\includegraphics[width=0.85\linewidth,]{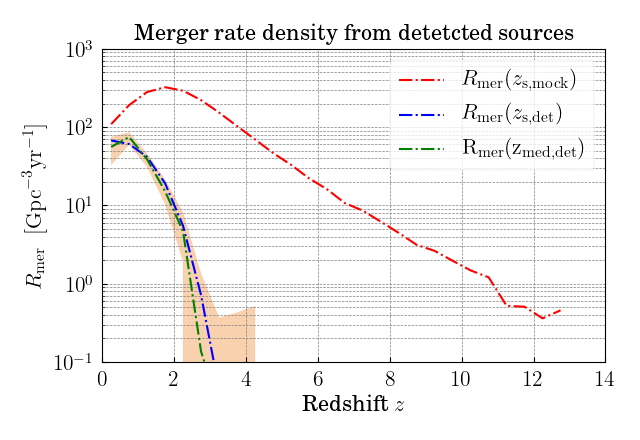}}
	\end{minipage}
	\begin{minipage}{0.50\linewidth}
		\centerline{\includegraphics[width=0.85\linewidth]{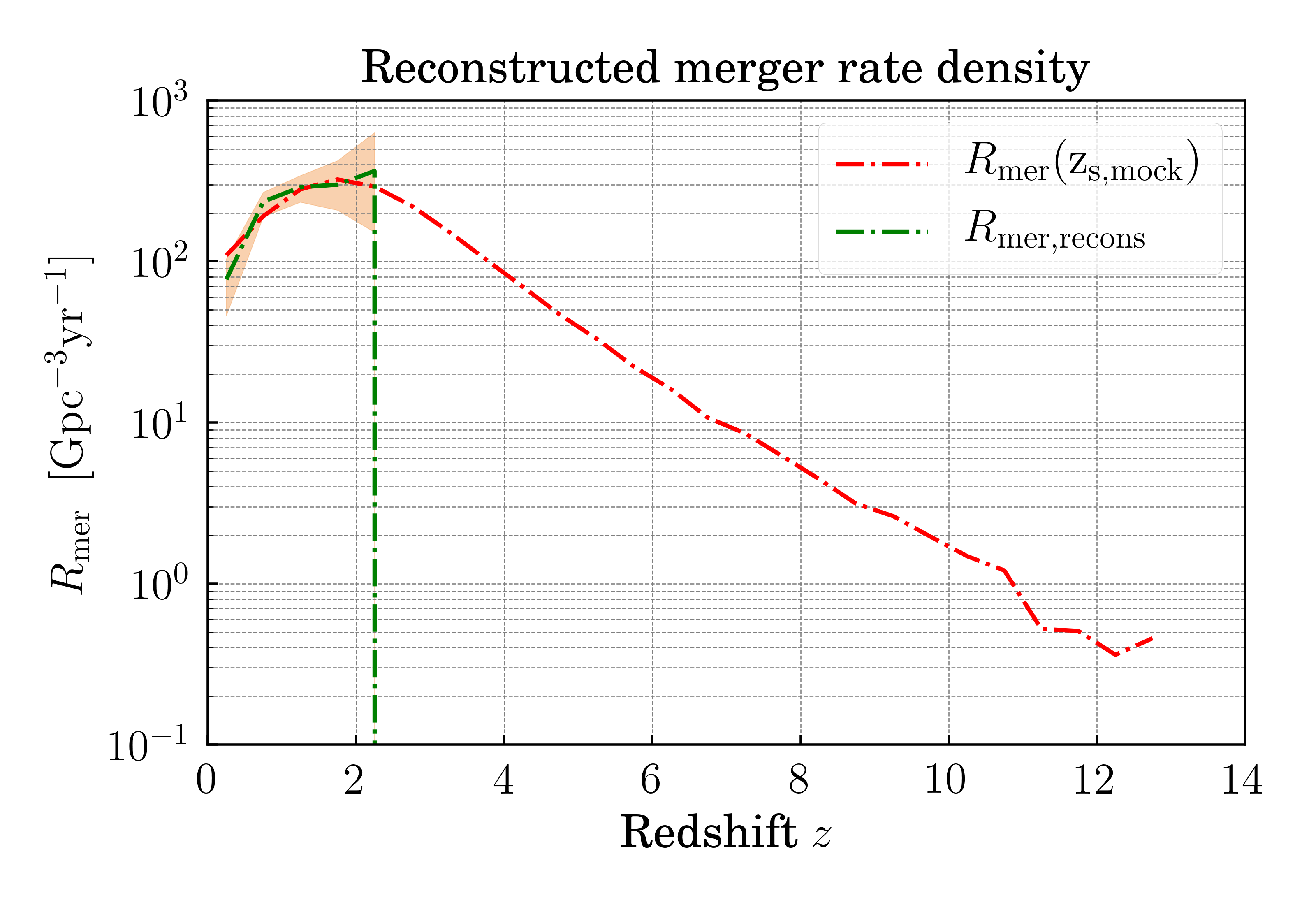}}
	\end{minipage}
	\caption[]{$R_{\rm mer}(z_{\rm s, mock})$ is with true source redshift in the whole mock population, $R_{\rm mer}(z_{\rm s, det})$ is with true sources redshift of only detected sources, and $R_{\rm mer}(z_{\rm med, det})$ is with the median of the estimated posterior distribution of the redshift of detected sources. }
	\label{fig:rate}
	\begin{minipage}{\linewidth}
		\centerline{\includegraphics[width=0.55\linewidth]{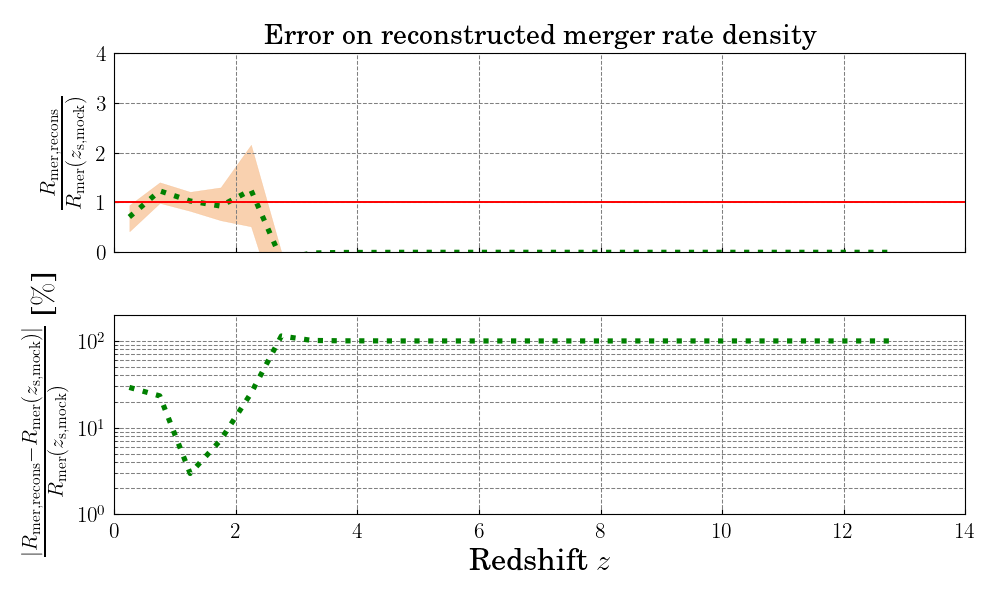}}
	\end{minipage}
	\caption[]{Error on the $R_{\rm mer, recons}$ estimate.}
	\label{fig:rate_err}
\end{figure}

\section{Estimation of the merger rate density}
For a population of BNS, if $N_{yr}$ is the number of expected mergers in a year, then the expected time taken for  $N_{mock}$ number of simulated BNSs in the mock population to merge is $T_{\rm mock} = \frac{N_{\rm mock}}{N_{\rm yr}}$ yr. Then the merger rate density $R_{\rm mer}$ for a given population, calculated for the time $T_{\rm mock}$ is:

\be\label{eq:merger_rate_den}
	R_{\rm mer}(z_i, z_{i+1}) = \frac{1+z_{i+1}}{\int^{z_{i+1}}_{z_i} \frac{dV}{dz}dz}\left( \frac{N_{(z_i,z_{i+1})}}{T_{\rm mock}}\right),
\ee
where $N_{(z_i,z_{i+1})}$ is the number of mergers in a redshift bin $[z_i:z_{i+1}]$. 
From Equation (\ref{eq:merger_rate_den}), we obtain three sets of merger rate densities for our BNSs population: (i) $R_{\rm mer}(z_{\rm s, mock})$, (ii) $R_{\rm mer}(z_{\rm s, det})$, and (iii) $R_{\rm mer}(z_{\rm med, det})$. These merger rate densities are shown in Figure~\ref{fig:rate}.

Figure~\ref{fig:rate}, left panel, shows that, $R_{\rm mer}(z_{\rm med, det}) \approx R_{\rm mer}(z_{\rm s, det})$, but $R_{\rm mer}(z_{\rm med, det}) \ll R_{\rm mer}(z_{\rm s, mock})$. This is because a large fraction of the BNSs do not cross the detection threshold we have chosen. In our mock population, less than 1/10 of the binaries were detected. In order to estimate the detection efficiency we assume that the detected population of the compact binaries truly represents the redshift and chirp mass distribution of the whole population. We construct a secondary mock population, denoted with subscript `sec', from the detected sources, assuming that the distributions of the chirp mass and redshift are proportional to $\mathcal{M}_{\rm med,det}$ and $z_{\rm med, det}$, that is, $p(\mathcal{M}_{\rm sec})\propto p(\mathcal{M}_{\rm med,det})$ and $p(z_{\rm sec}) \propto p(z_{\rm med,det})$. We assume that the mass ratio $q_{\rm sec}$ is uniformly distributed in the range [0,1] with a constraint on total mass $M_{\rm sec}$ such that $(M_{\rm med,det})_{\rm min} \leq M_{\rm sec} \leq (M_{\rm med,det})_{\rm max}$, where $M_{\rm sec}$ is defined as	$M_{\rm sec} = \mathcal{M}_{\rm sec} \left[\frac{q_{\rm sec}}{(1+q_{\rm sec})^2} \right]^{-3/5}$ (see details in Singh {\it et al}~\cite{2024A&A...681A..56S}). We then, use this secondary population to estimate the detection efficiency. The detection efficiency $\mathcal{D}$ as a function of redshift is then defined as $\mathcal{D}(z_i, z_{i+1}) = \left[\frac{N_{\rm sec, det}}{N_{\rm sec}}\right]_{(z_i,z_{i+1})}$, where $[N_{\rm sec}]_{(z_i,z_{i+1})}$ is the number of mergers in the secondary mock population in the redshift bin $(z_i,z_{i+1})$ and $[N_{\rm sec, det}]_{(z_i,z_{i+1})}$ is the mergers in this bin that crossed the detection threshold. Using this detection efficiency we reconstruct the merger rate density $R_{\rm mer, recon}$ which is then given as:

\be\label{recons_mer}
	R_{\rm mer, recons}(z_i, z_{i+1}) = \left[\frac{R_{\rm mer}(z_{\rm med, det})}{\mathcal{D}}\right]_{(z_i, z_{i+1})} .
\ee

Figure~\ref{fig:rate}, right panel, shows the reconstructed merger rate density. The red line is the true merger rate density, while the green line show the merger rate density calculated using Equation (\ref{recons_mer}). The shaded region represents the Poisson error for 0.54 month of observation time. To quantify the accuracy of the reconstruction, we show the relative error on the merger rate density $\frac{|R_{\rm mer}(z_{\rm s, mock}) - R_{\rm mer, recon} |}{R_{\rm mer}(z_{\rm s, mock})}$ as a function of redshift in Figure~\ref{fig:rate_err}.

\section{Conclusion}
In our earlier work~\cite{2024A&A...681A..56S} we developed a population independent method to reconstruct the merger rates of coalescing compact binaries with ET using information from the full inspiral of a coalescing BNS signal to localise and then constrain the intrinsic parameters of the binary. We formulated a scheme to estimate accurate detection efficiency and showed how to estimate the true merger rate density of the underlying population. In the current work, we used the algorithm developed earlier to explore the capability of ET as a single instrument to constrain the merger rate density of the BNSs as a function of redshift. We show that for a population of BNSs with $R_{mer} \sim 100 (300)$ $\rm Gpc^{-3} yr^{-1}$ at $z\sim 0(2)$, we can reconstruct the merger rate density uptil $z \sim 2$ , with a relative error of $12\%$ at ($z \sim 2$), despite the loss in detection of the bulk of the BNS population.

\section*{Acknowledgments}

This work was supported by the Universitat de les Illes Balears (UIB); the Spanish Agencia Estatal de Investigación grants CNS2022-135440, PID2022-138626NB-
I00, RED2022-134204-E, RED2022-134411-T, funded by MICIU/AEI/10.13039/501100011033, and the European Union NextGenerationEU/PRTR, and the ERDF/EU; and the Comunitat Autònoma de les Illes Balears through the Conselleria d’Educació i Universitats with funds from the European Union - NextGeneration EU/PRTR-C17.I1 (SINCO2022/6719) and from the European Union - European Regional Development Fund (ERDF) (SINCO2022/18146).TB acknowledges support of the NCN grant 2023/49/B/ST9/02777.

\section*{References}

\bibliography{ET_BNS}

\end{document}